\documentclass{raa}
\usepackage{graphicx,times}
\usepackage{natbib}
\usepackage{amssymb,amsmath}
\bibpunct{(}{)}{;}{a}{}{,}

\usepackage[a4paper=true,pagebackref=true]{hyperref}
\hypersetup{pdftitle = The title of my PDF, pdfauthor = My name, pdfsubject= The subject, pdfkeywords = keyword1 keyword2 keyword3}
\hypersetup{colorlinks = true, linkcolor = green, anchorcolor = red, citecolor = blue, filecolor = red, pagecolor = red, urlcolor = red}

\begin{document}

   \title{Anisotropic power spectrum and the observed low-$l$ power in PLANCK CMB data
$^*$
\footnotetext{\small $*$ Supported by the National Natural Science Foundation of China.}
}

 \volnopage{ {\bf 2017} Vol.\ {\bf X} No. {\bf XX}, 000--000}
   \setcounter{page}{1}

   \author{Zhe Chang\inst{1,2}, Pranati K. Rath\inst{1,3}, Yu Sang\inst{1,2}, Dong Zhao\inst{1,2}}

   \institute{ Institute of High Energy Physics, Chinese Academy of Sciences, Beijing 100049, China\\
    \and
     University of Chinese Academy of Sciences, Beijing 100049, China\\
        \and
         Theoretical Physics Center for Science Facilities, Chinese Academy of Sciences, Beijing 100049, China;{\it pranati@ihep.ac.cn} \\
\vs \no
{\small }
}

\abstract{
In this work, we study a direction dependent power spectrum in anisotropic
Finsler space-time. We use this direction dependent power spectrum to address the low-$l$ power observed in WMAP and PLANCK data.
The angular power spectrum of the temperature fluctuations has a lower amplitude in comparison to the $\Lambda$CDM model
in the multipole range $l = 2-40$. Our theoretical model gives a correction to the isotropic angular power spectrum $C^{TT}_l$
due to the breaking of the rotational invariance of the primordial power spectrum.
We estimate best-fit model parameters along with the six $\Lambda$CDM cosmological parameters using PLANCK likelihood code in CosmoMC software.
We see that this modified angular power spectrum fits the CMB temperature data in the multipole range $l=2-10$ to a good extent but
fails for the whole multipole range $l = 2-40$.
}

\authorrunning{Zhe Chang et al.}
\titlerunning{Anisotropic power spectrum and the observed low-$l$ power in PLANCK CMB data}

\maketitle

%
\section{Introduction}
\label{sect:intro}
The standard Lambda cold dark matter ($\Lambda$CDM) cosmological model predicted by the inflationary scenario at the
very early Universe is impressively successful in explaining the observed Cosmic Microwave Background (CMB) data.
However, a set of CMB observations which are not statistically consistent with the $\Lambda$CDM model has been observed in both WMAP and PLANCK CMB data.
These observations include alignment of CMB quadrupole and octopole
(\citealt{Costa2004,Copi2004,Ralston2004,Kate2005,Abramo2006a,Abramo2006b,Copi2015a}),
lack of power at large scale up to $l \le 40$ (\citealt{Jing1994,Bennett2011,Plancklik2014,Iqbal2015}),
the lack of large angular correlations on angular scales larger than $60^o$ (\citealt{Spergel2003,Copi2009, Copi2015b})
and hemispherical power asymmetry (\citealt{Eriksen2004a,Eriksen2007a,Erickcek2008a,Erickcek2008b,Hansen2009,Hanson2009,Groeneboom2010,Hoftuft2009,Planckiso2016,Pranati2013b,Pranati2015a,Pranati2015b}).
The CMB observations also suggest
parity asymmetry (\citealt{Kim2010b,Kim2010c,Gruppuso2011,Kim2011,Aluri2012a,BenDavid2012,Zhao2014,Shiraishi2015,Aluri2017})
and a cold spot in southern hemisphere (\citealt{cruz2005,Cruz2006,Cruz2008,Vielva2010,Lim2012}).
The significance of these observations has been motivated many theorists to study different theoretical models.
Hence there exists a number of theoretical models based on anisotropic space-times (\citealt{Berera2004,Kahniashvili2008,ACW2007,Chang2013c})
and inhomogeneous universe (\citealt{Moffat2005}).The theoretical models violating the rotational invariance lead to a direction dependency in the primordial power spectrum
(\citealt{ACW2007,Goldwirth1990, Emir2007, Pontzen2007, Pereira2007,Pullen2007,Campanelli2009,Donoghue2009,Watanabe2009,Chang2013c}).

The primordial power spectrum $P(k)$ defined as the two-point correlation function of the primordial
density perturbation $\delta(\vec k)$ can be written as
\begin{equation}
\langle{\delta(\vec k)\delta^*(\vec k')}\rangle = (2\pi)^3 \delta^3(\vec k-\vec k') P(k)\, .
\label{eq:pk}
\end{equation}
The Dirac delta function of Eq. (\ref{eq:pk}) ensures that the modes with
different wave numbers are not coupled with each other which is the consequence of the translational invariance.
In the standard ${\Lambda}$CDM model which refers to the homogeneous and isotropic FRW metric, the
fluctuations are statistically isotropic and the primordial power spectrum $P(k)$ depends only
on the magnitude of the wave vector $\vec k$. Hence the primordial power spectrum is rotationally invariant and
one can write the primordial power spectrum, $P(k)$ as
\begin{equation}
P(k) = A_{s}\Bigg(\frac{k}{k_{c}}\Bigg)^{n_{s} - 1} \,
\end{equation}
where $n_s$ is the spectral index, $A_s$ is the spectral amplitude and $k_c$ is the scalar pivot.
In this case, the spherical harmonic coefficient $a^{T}_{lm}$ of the temperature fluctuation obeys the statistical isotropy and
hence the two-point correlation of $a^T_{lm}$ can be written as
\begin{equation}
\left\langle{a^{T}_{lm}a^{T*}_{l^{\prime}m^{\prime}}}\right\rangle = C^{TT}_l\delta_{ll'}\delta_{mm'} \,
\end{equation}
where $C^{TT}_l$ is the angular power spectrum encoding all the information of the CMB temperature fluctuations.

But in case of an anisotropic space-time which breaks the rotational invariance of the power spectrum, the spherical harmonic coefficient $a^{T}_{lm}$
no longer follow the statistical isotropy and the two-point correlation function of $a^{T}_{lm}$ give rise
to off-diagonal correlation between multipole moments.
The off-diagonal correlations encode all the crucial information regarding the anisotropic model.
Hence one can write
\begin{equation}
\left\langle{a^{T}_{lm}a^{T*}_{l^{\prime}m^{\prime}}}\right\rangle \equiv C^{TT}_{ll'mm'}\,
\label{eq:almaniso}
\end{equation}

In WMAP and PLANCK data, it has been observed that the temperature angular power spectrum,
$C^{TT}_l$, at low-$l$ ($l \le 40$) have a lower amplitude than the $\Lambda$CDM model
(\citealt{Bennett2011,Planck2014XVI,Planck2014XV,Planck2016liki}).
In Ref.~\cite{Hazra2014}, the authors also studied the consistency of the $\Lambda$CDM model with the Planck data
and claimed that the data has lack of power at both
high and low $l$ multipoles. This issue has been studied extensively by many theorists in the inflationary framework (\citealt{Contaldi2003,Boyanovsky2006,Cicoli2014,Das2014}).
In this paper, we try to relate the direction dependent power spectrum with the lack of power at large scale and find out the best-fit model parameters.

The paper is organized as follows. In section \ref{sec:ps}, we review briefly about Finsler space-time and a direction dependent power spectrum in this space-time.
Then we implement this power spectrum to study the lack of power at large scale. To study its effect on the angular power spectrum $C^{TT}_l$,
we perform Monte Carlo Markov Chain (MCMC) analysis using Planck data. In section \ref{sec:mcmc}, we present the results of the MCMC analysis.
In section \ref{sec:con}, we summarize our work.

\section{Anisotropic Model}
\label{sec:ps}
Here we briefly review an anisotropic space-time in the framework of Finsler geometry (\citealt{Chang2009,Chang2013cosmo,Chang2013c,Li2015}).
In Refs.~(\cite{Chang2013c,Li2015}), the authors have studied the anisotropic inflation taking Finslerian background spacetime.
The Finsler spacetime has fewer symmetries than the Riemann symmetry and hence is a suitable candidate to study the anisotropy
observations.
The counterparts of special relativity (\citealt{Gibbons2007,Chang2008,Chang2012b}) commonly known as very special relativity (VSR) (\citealt{Coleman1997,Coleman1999,Cohen2006})
have connections with the Finsler geometry (\citealt{Bao2000}) which is
generalized from Riemann geometry by removing the quadratic restriction.  In order to investigate these counterparts, one should study the
inertial frames and symmetry in Finsler spacetime. The symmetry of spacetime is described by investigating the Killing vectors (\citealt{Li2012symmetry}).
Finsler geometry is defined on the tangent bundle with proper length, $s$, as
\begin{equation}
s = \int_a^b F(x,y)ds \,
\end{equation}
where $x$ and $y\equiv {dx/ds}$ are the positions and the velocity respectively.
The integrand $F(x,y)$ which is known as the Finsler structure is the basis of Finsler geometry.
This is a smooth and positive function on the tangent bundle of a manifold $M$.
For any $\lambda > 0$, Finsler structure $F$ obeys
\begin{equation}
 F(x,\lambda y)=\lambda F(x,y).
\end{equation}
The Finsler metric is given by the second derivative of $F^2$ with respect to velocity $y$ as
\begin{equation}
g_{\mu\nu}= \frac{\partial}{\partial y^\mu}\frac{\partial}{\partial y^\nu}\left(\frac{1}{2}F^2\right) \,
\label{eq:FSmetric}
\end{equation}
where the spatial indices of $\mu$ and $\nu$ run from $1$ to $3$ and the temporal index is $0$.
A Finsler metric is said to be locally Minkowskian if at every point, there exist a local coordinate system in which the Finsler
structure $F$ is independent of the position $x$, i.e, $F = F(y)$. This is known as the flat Finsler space-time.
The flat Finsler space-time can be used to test the Lorentz invariance through the modified dispersion relation.
The geodesic equations in Finsler space-time can be given by the first order variation of Finslerian length as (\citealt{Li2017cpc})
\begin{equation}
\frac{d^2x^\mu}{d\tau^2}+2G^\mu=0,
\end{equation}
where the geodesic spray coefficient $G^\mu$ is given as
\begin{equation}
G^\mu=\frac{1}{4}g^{\mu\nu}\left(\frac{\partial^2F^2}{\partial x^\lambda\partial y^\nu}y^\lambda-\frac{\partial F^2}{\partial x^\nu}\right).
\end{equation}
The coefficient $G^\mu$ vanishes in the locally Minkowski space.

The observed CMB anomalies may be related to a special case of Finsler space-time known as Randers-Finsler space-time.
The Randers space (\citealt{Randers1941}) involves a vector field which may influence the anisotropic evolution of the early universe.
The structure is given by
\begin{equation}
F^2=y^ty^t-a^2(t)F^2_{Ra}.
\end{equation}
Here $F^2_{Ra}$ is the structure of Randers space, and
\begin{equation}
F^2_{Ra}(x,y)=\alpha(x,y)+\beta(x,y),
\label{eq:Fsquare}
\end{equation}
where $\alpha(x,y)=\sqrt{\tilde{a}_{\mu\nu}(x)y^\mu y^\nu}$ is a Riemann structure with metric $\tilde{a}_{\mu\nu}$,
and $\beta(x,y)=\tilde{b}_{\mu}(x)y^\mu$ is a 1-form. This vector induces the anisotropic properties in the Randers space.
Here, $\tilde{a}_{\mu\nu}$ can be taken as the flat FRW metric,
and $\tilde{b}_{\mu}$ has only the temporal component, i.e., $\tilde{b}_{\mu}=(B(z),0,0,0)$, where $B(z)$ depends on the third spatial coordinate $z$.
Finsler metric will be reduced to FRW metric if $B(z)\rightarrow0$.
The 1-form $\beta(x,y)$ is relevant to a vector field, which will give a privileged axis in the space-time.

To investigate the Killing vector, one should discuss the isometric transformation under an infinitesimal coordinate transformation.
The isometric transformation for $x$ and $y$ are given as,
\begin{eqnarray}
\bar x^{\mu} &=& x^{\mu} + \epsilon V^{\mu}\\
\bar y^{\mu} &=& y^{\mu} + \epsilon \frac{\partial V^{\mu}}{\partial x^{\nu}} y^{\nu} \,.
\end{eqnarray}
In the first order of $\epsilon$, the Finsler structure is,
\begin{equation}
\bar F(\bar x, \bar y) = \bar F(x, y) + \epsilon V^{\mu} \frac{\partial F}{\partial x^{\mu}}
+ \epsilon y^{\nu} \frac{\partial V^{\mu}}{\partial x^{\nu}} \frac{\partial F}{\partial y^{\mu}} \,.
\end{equation}
The Finsler structure is called isometry if and only if $F(x,y) = \bar F(x,y)$. Hence one can obtain the Killing equation in Finsler space as
\begin{eqnarray}
K_V(F) &\equiv& V^{\mu} \frac{\partial F}{\partial x^{\mu}} + y^{\nu} \frac{\partial V^{\mu}}{\partial x^{\nu}} \frac{\partial F}{\partial x^{\mu}} = 0
\label{eq:killeq}
\end{eqnarray}
Using Eq. (\ref{eq:Fsquare}), one can see that the number of independent Killing Vectors in Randers-Finselr space-time is less than Riemannian space-time.

The speed of light is direction dependent in Finsler space-time. Along the radial direction, it can be derived as (\citealt{Li2010,Li2014,Li2015epjc})
\begin{equation}
c_r=\frac{1}{1+Bcos\theta},
\label{eq:vellight}
\end{equation}
where $\theta$ is the angle along the $z$-axis. Hence the redshift in Finsler space-time is
 \begin{equation}
1+z=\frac{1+Bcos\theta}{a}.
\label{eq:redshift}
\end{equation}
The variation of speed light Eq. (\ref{eq:vellight}) gives a variation of the fine-structure constant which is a dipolar distribution.
This dipole distribution of the fine structure constant is in agreement with the observations of the quasar absorption spectra (\citealt{Webb2011,King2012,Chang2012a}).
Using Eq. (\ref{eq:redshift}), the luminosity distance in Finslerian universe is given as
\begin{equation}
d_{L} = (1+z)r = \frac{1+z}{H_0}\int_0^z \frac{dz}{\sqrt{\Omega_{m0}(1+z)^{3}(1-3B\cos{\theta})+1-\Omega_{m0}}}\,
\end{equation}
where the radial distance $r=\sqrt{x^2+y^2+z^2}$.

In the standard cosmological model, the power spectrum is derived in isotropic space-time.
However, if there exists a privileged direction in space-time, the early evolution of universe will have different behaviours.
This anisotropic space-time at the early stage of inflation breaks the rotational invariance of the primordial power spectrum and
leads to a direction dependent power spectrum.
Taking Randers space-time with a weak vector field, i.e., $|\tilde{b}_{\mu}|<<1$, as the background space-time of inflation and
solving the equation of motion of the inflaton field, one can obtain a direction dependent power spectrum of the form
\begin{equation}
P'(k) = P_{iso}(k)\Bigg(1+i A(k) \left(\hat k \cdot \hat n\right) + B(k)\left(\hat k \cdot \hat n\right)^2\Bigg) ,
\label{eq:ps}
\end{equation}
where $P_{iso}(k)$ denotes the isotropic power spectrum, $A(k)$ and $B(k)$ are some arbitrary functions of wave number $k$.
The function $A(k)$ and $B(k)$ encode the amplitude of dipolar and quadrupolar modulation to the isotropic power spectrum.
We restrict ourselves to second order correction of the isotropic primordial power spectrum as
the next higher order terms in $(\hat k \cdot \hat n)$ will be suppressed by the magnitude
of the small vector. The breaking of the rotational invariance of the primordial power spectrum leads to the non-vanishing
correlations between different multipole moments that would normally vanish.
The same type of direction dependent power spectrum in the leading order of
$(\hat k \cdot \hat n)$ has obtained in the Refs.~\cite{Pranati2015a,Pranati2015b,Kothari2016,Shamik2016,Zibin2017,Chang2013c,Li2015}
to address the hemispherical power asymmetry successfully.
The authors in Refs.~\cite{Pranati2015a,Pranati2015b,Kothari2016,Shamik2016} constrained the amplitude in the multipole range $l=2-64$
with a $3\sigma$ confidence level(CL) using PLANCK data. The amplitude for the quadrupolar modulation $B(k)$ has been constrained by
the Refs.~\cite{Kim2013,Plancknon2016} and they found it to be an order of $10^{-2}$.
Here we are not giving any remark on the quadrupolar modulation constraint and focus only on the
correction to the isotropic power spectrum due to the quadrupolar modulation in the power spectrum.

\section{Application on CMB data}
The temperature fluctuation in terms of primordial density fluctuations $\delta(k)$  can be written as
\begin{equation}
 \frac{\Delta T}{T_0}(\hat n)= \int d^{3}k\sum_{l}\frac{2l+1}{4\pi}(-i)^{l}P_{l}(\hat k\cdot \hat n)\delta(k) \Delta^{T}_{l}(k)\, ,
\label{eq:temp}
\end{equation}
where $P_{l}$ and $\Delta^{T}_{l}(k)$ are the Legendre polynomial and the transfer function of order $l$ respectively.
The transfer function helps in understanding the change in amplitude of the perturbation from an initial time to the current time.
Now using Eq. (\ref{eq:temp}) one can write the spherical harmonic coefficients $a^{T}_{lm}$ as
\begin{equation}
a^{T}_{lm} = \int d\Omega Y_{lm}^{*}(\hat n){\Delta T}(\hat n) \, ,
\end{equation}
and the two-point correlation function of $a^{T}_{lm}$ as
\begin{equation}
\langle{a^{T}_{lm}a^{T*}_{l'm'}}\rangle = {\langle{a^{T}_{lm}a^{T*}_{l'm'}}\rangle}_{iso}+{\langle{a^{T}_{lm}a^{T*}_{l'm'}}\rangle}_{aniso} \, ,
\end{equation}
where the first term gives the isotropic angular power spectrum $C^{TT}_l$
\begin{equation}
C^{TT}_l = \int_{0}^{\infty}k^{2}dkP_{iso}(k)({\Delta_l^{T}(k))^{2}} \, ,
\end{equation}
and the second term contains all the anisotropic terms. Following Eq. (\ref{eq:almaniso}), one can write the
anisotropic term as
\begin{equation}
C^{TT}_{ll'mm'} = {\langle{a_{lm}a^{*}_{l'm'}}\rangle}_{dm} + {\langle{a_{lm}a^{*}_{l'm'}}\rangle}_{qm} \, ,
\end{equation}
where the dipole modulation term is given as
\begin{equation}
{\langle{a_{lm}a^{*}_{l'm'}}\rangle}_{dm} = (-i)^{l-l'}\xi^{dm}_{lm;l'm'}\int_{0}^{\infty}k^{2}dkP_{iso}(k)A(k){\Delta^{T}_{l}(k)}{\Delta^{T}_{l'}(k)}\,.
\end{equation}
and the quadrupolar modulation term is given as
\begin{equation}
{\langle{a_{lm}a^{*}_{l'm'}}\rangle}_{qm} = (-i)^{l-l'}\xi^{qm}_{lm;l'm'}\int_{0}^{\infty}k^{2}dkP_{iso}(k)B(k){\Delta^{T}_{l}(k)}{\Delta^{T}_{l'}(k)}\,.
\end{equation}
Following Refs.~\cite{ACW2007,Pranati2013a}, we use the spherical components of the unit vector $n$ as
\begin{equation}
 n_{+} = -\left(\frac{n_{x}-in_{y}}{\sqrt 2}\right),n_{-} = \left(\frac{n_{x}+in_{y}}{\sqrt 2}\right),n_0 = n_{z}\ .
\end{equation}
The geometrical factor $\xi^{dm}_{lm;l'm'}$ of dipolar modulation term is defined as
\begin{equation}
 \xi^{dm}_{lm;l'm'}  =  n_{+}\xi^{dm+}_{lm;l'm'}+n_{-}\xi^{dm-}_{lm;l'm'}+n_{0}\xi^{dm0}_{lm;l'm'} \, ,
\end{equation}
which gives the correlation between multipoles moments differ by $\Delta l = 1$ and it has no effect on the isotropic angular power spectrum $C^{TT}_l$.
Hence by taking the preferred axis along $z$-axis, the coefficients of $\xi^{dm}_{lm;l'm'}$ can be given as
\begin{equation}
\xi^{dm0}_{lm;l'm'} = \delta_{m',m}\left[\sqrt{\frac{(l-m+1)(l+m+1)}{(2l+1)(2l+3)}}\delta_{l',l+1} + \sqrt{\frac{(l-m)(l+m)}{(2l+1)(2l-1)}}\delta_{l',l-1}\right] \, .
\end{equation}
 This term successfully explained the observed hemispherical power asymmetry (\citealt{Pranati2015a,Kothari2016,Shamik2016,Chang2013c,Li2015}).

Next, we will discuss the quadrupolar modulation term in the power spectrum.
The geometrical factor $\xi^{qm}_{lm;l'm'}$ of the quadrupolar modulation term is given as
\begin{eqnarray}
\xi^{qm}_{lm;l'm'} &=& n^{2}_{+}\xi^{qm++}_{lm;l'm'}+n^{2}_{-}\xi^{qm--}_{lm;l'm'}+2n_{+}n_{-}\xi^{qm+-}_{lm;l'm'}+2n_{+}n_{0}\xi^{qm+0}_{lm;l'm'}\nonumber\\
&&+2n_{-}n_{0}\xi^{qm-0}_{lm;l'm'}+n^{2}_{0}\xi^{qm00}_{lm;l'm'} \, .
\end{eqnarray}
This term contains all the correlation between multipoles differ by $\Delta l = 2$ and $\Delta l = 0$.
Hence the isotropic angular power spectrum $C^{TT}_l$ changes if we consider the coefficients with $\Delta l = 0$.
The coefficients of $\xi^{qm}_{lm;l'm'}$ for $l' = l$ and $m' = m$ are
\begin{eqnarray}
 \xi^{qm+-}_{lm;l'm'} &=& - \delta_{m',m} \, \frac{(l^2+m^2+l-1)}{(2l-1)(2l+3)} \\
 \xi^{qm00}_{lm;l'm'} &=& \delta_{m,m'} \, \frac{(2l^{2}+2l-2m^{2}-1)}{(2l-1)(2l+3)}
\end{eqnarray}
By setting the preferred direction along the $z-axis$, only $\xi^{qm00}_{lm;l'm'}$ will contribute to $C^{TT}_l$.
This correction depends on the anisotropic power spectrum $B(k)$. Hence to estimate its effect on $C^{TT}_l$,
we parameterize the anisotropic power spectrum $B(k)$. We try two forms of the anisotropic power spectrum $B(k)$, first one is the
power law form and the second one is the exponential form.
The Power law form of anisotropic power spectrum is given as
\begin{equation}
B(k) = -B_0 \Bigg(\frac{k}{k_c}\Bigg)^{-\alpha} \,
\label{eq:pl}
\end{equation}
and the exponential form of the anisotropic power spectrum is given as
\begin{equation}
B(k) = B_0 \exp \Bigg[-\Bigg(\frac{k}{k_c}\Bigg)^{\alpha}\Bigg] \,
\label{eq:expf}
\end{equation}
where $B_0$ and $\alpha$ are the amplitude and the spectral index of the anisotropic term. In the next section, we will use both the forms of $B(k)$ and
estimate the theoretical model parameters $B_0$ and $\alpha$ in addition to six cosmological parameters using CosmoMC software.

\section{Analysis and Results}
\label{sec:mcmc}
For our analysis, we use publicly available CosmoMC software (\citealt{Lewis2002}) which consists of Fortran and python codes.
CosmoMC uses CAMB (\citealt{Lewis2000}) code to compute the theoretical angular power spectrum and
uses Markov-Chain Monte-Carlo (MCMC) to compute the best-fit cosmological parameters.
To get the best-fit parameters using likelihood, we use the PLANCK likelihood code (PLC/clik) provided by
PLANCK team with CosmoMC software (\citealt{Planck2014XV}).
The PLANCK likelihood code uses COMMANDER at low-$l$ ($l=2-49$) and CamSpec code at high-$l$ ($l=50-2500$).
The inputs to the CosmoMC are the central values and the flat priors of the various model parameters.
We use CosmoMC's python scripts and getdist to analyze the generated chains from the MCMC analysis and to produce the required plots.

We modify the required CAMB and CosmoMC code using Eq. (\ref{eq:ps}) for our analysis. We use Eq. (\ref{eq:pl}) and (\ref{eq:expf}) for the
anisotropic part of the Eq. (\ref{eq:ps}). We use flat priors for the model parameter $B_0$ and $\alpha$ in addition to the six $\Lambda$CDM parameters
as the input to the MCMC analysis. The list of parameters and their prior ranges are listed in Table \ref{tab:prior1} and \ref{tab:prior2}.
We first check for the power law case of the anisotropic power spectrum Eq. (\ref{eq:pl}) and then move to the exponential form Eq. (\ref{eq:expf}).
For the power law case, we first run for both the parameters and get negative $C_l$ error in the CosmoMC for some range of $B_0$ and $\alpha$.
The reason for getting negative $C_l$ for those parameters is due to the larger value of the anisotropic term compared to the isotropic power spectrum.
This is not acceptable at all. Hence we try by fixing one of these two parameters.
We first fix $\alpha$ to different values and search the best-fit value of $B_0$.
Especially, by fixing $\alpha$ to $0.5$, we found the best-fit value of $B_0$ is $0.0342\pm0.0396$ which can explain the lack of power in low-$l$.
But as we see the error in $B_0$ is larger
than the best-fit value, we can not use this result. So we next try by fixing $B_0$ and allowing $\alpha$ to run in the range $[0,0.8]$.
We find that for $B_0=0.04$ and $\alpha = 0.4556 \pm 0.2158$, the theoretical model is able to explain the lack of power up to $l=10$ to a good extent.
This fitting is not as good as we wanted. Nonetheless, we listed all the best-fit parameter in Table \ref{tab:para}.

\begin{figure}[t!]
\centering
\includegraphics[width=0.80\textwidth]{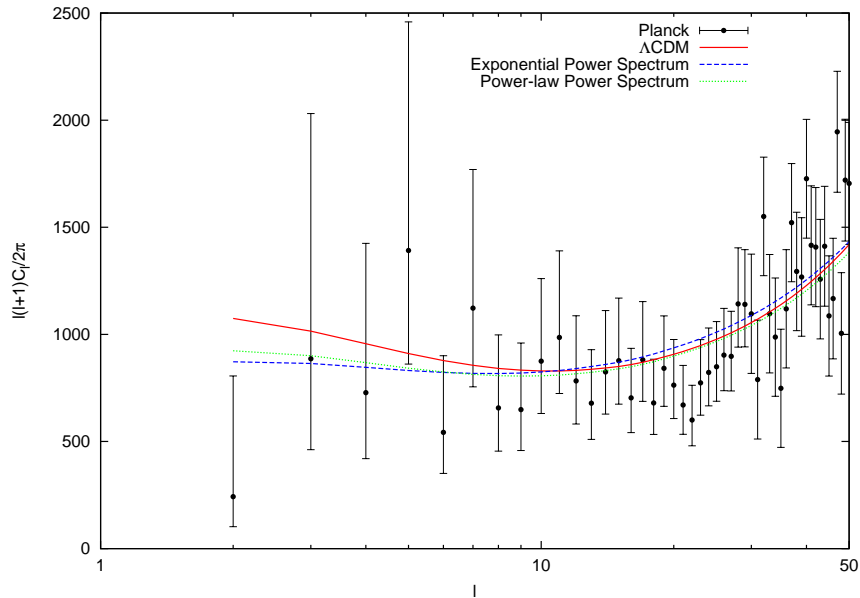}
\caption{The point with error bar represents the PLANCK 2015 temperature power spectrum for $l =2-50$. 
The red solid line represents the $\Lambda$CDM power spectrum and the green-dotted line
represents the theoretical power spectrum for the power law case with
the best-fit parameters $(B_0, \alpha) = (0.04, 0.4556 \pm 0.2158)$. The blue-dottted line represnts the theoretical power spectrum for
the exponential form with the best-fit parameters $(B_0, \alpha) = (0.4229 \pm 0.1134, 4.2889 \pm 1.2173)$. }
\label{fig:cl}
\end{figure}

\begin{figure}[t!]
\centering
\includegraphics[width=0.80\textwidth]{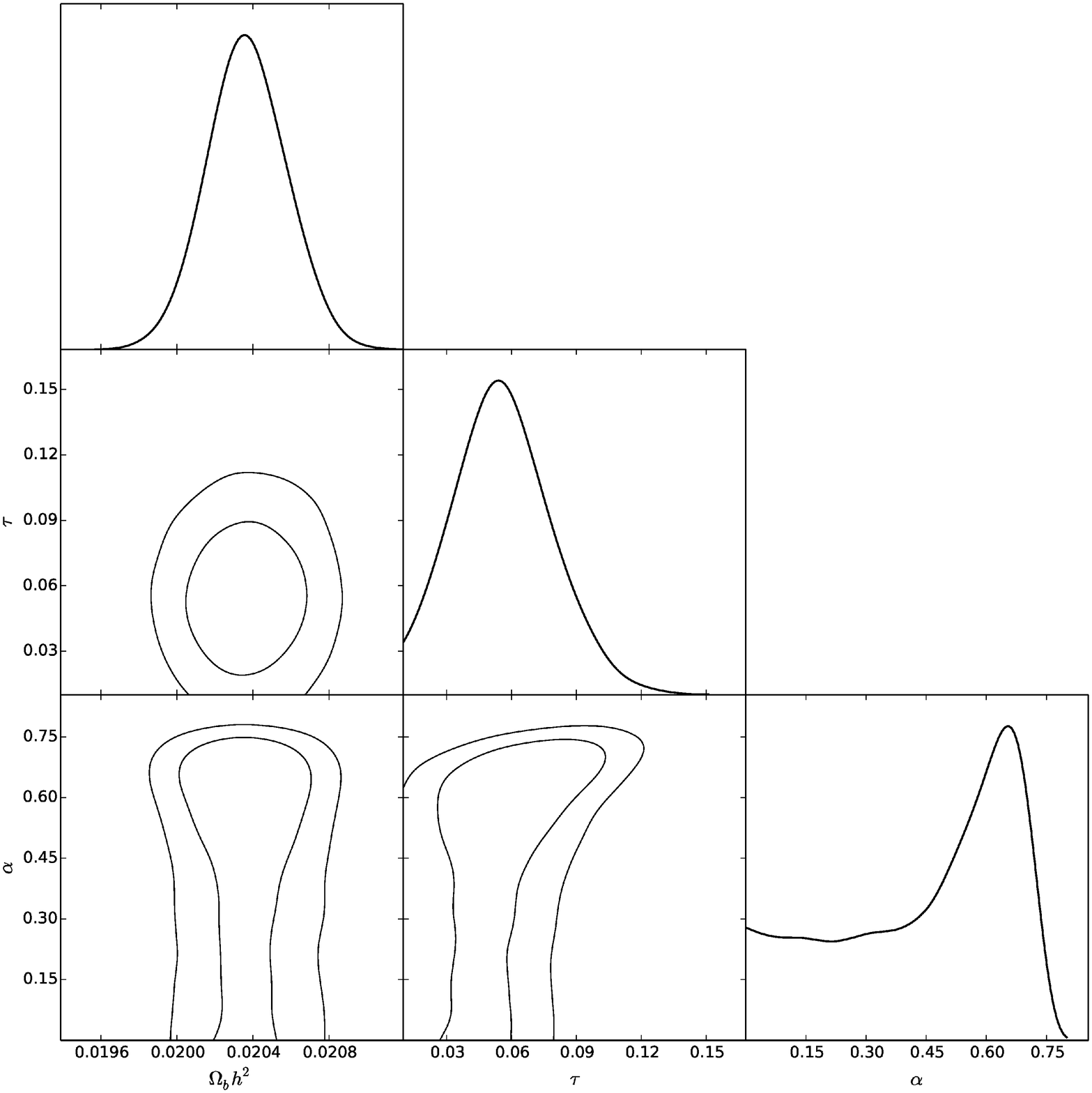}
\caption{The parameter plot for the power-law form of the anisotropic term in the power spectrum}
\label{fig:contour1}
\end{figure}

\begin{figure}[t!]
\centering
\includegraphics[width=0.80\textwidth]{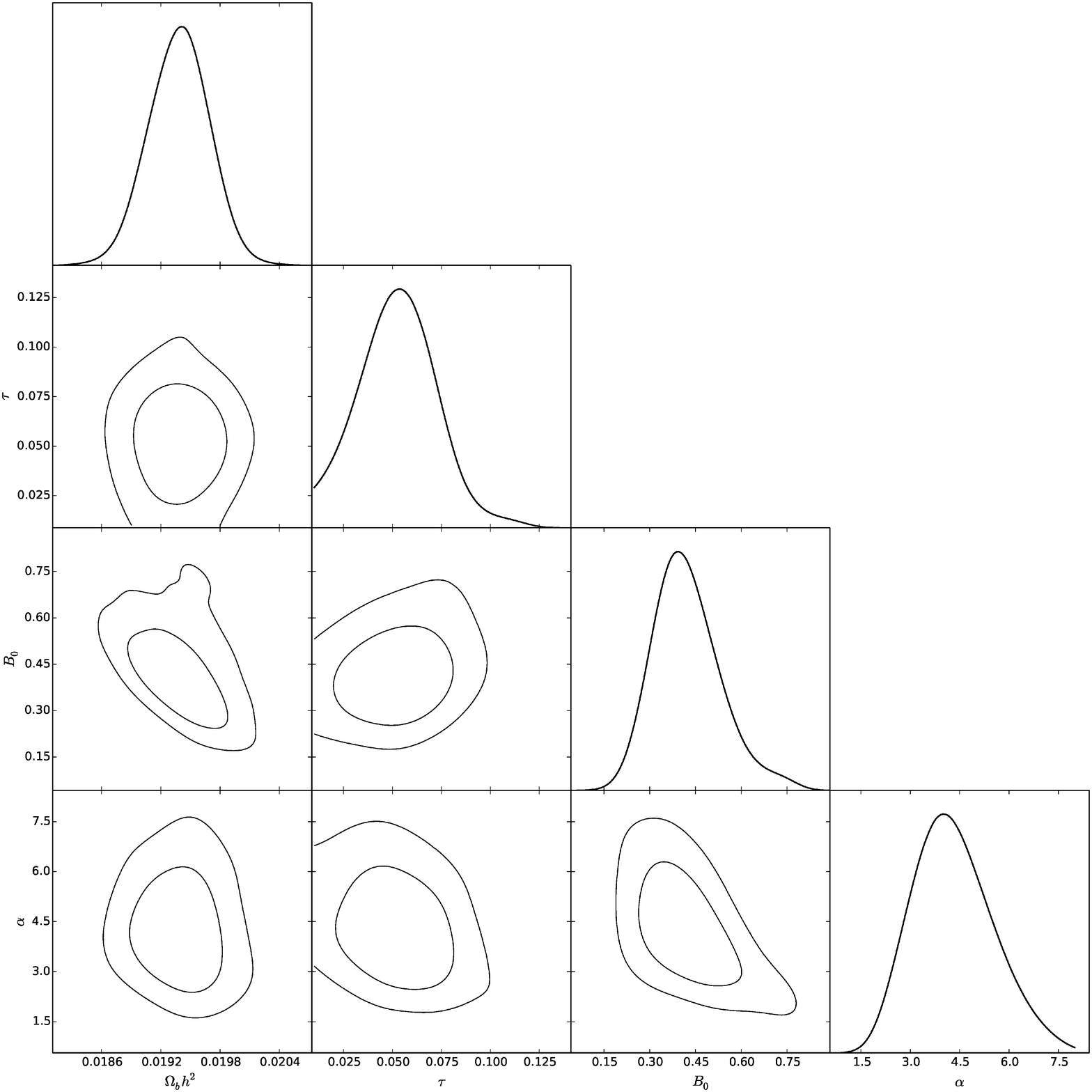}
\caption{The parameter plot for the exponential form of the anisotropic term in the power spectrum}
\label{fig:contour2}
\end{figure}

Next, we try for the exponential form of the anisotropic power spectrum. In this case, we allow both the model parameters to vary.
We choose to run the parameters in the range $\alpha = [0, 8]$ and $B_0 = [-1, 1]$ respectively.
By searching the best-fit value in the chosen wide range, we find the best-fit values as $\alpha=4.2889 \pm 1.2173 $ and $B_0= 0.4229 \pm 0.1134$.
The best-fit parameter value from the MCMC analysis are given
in Table \ref{tab:para}.
In Fig. \ref{fig:cl}, we plot the PLANCK 2015 temperature power spectrum
along with the best-fit theoretical power spectrum obtained from $\Lambda$CDM and from our theoretical model.
In this Figure, the power law and the exponential form of the anisotropic power spectrum takes $(B_0, \alpha)=(0.04, 0.4556 \pm 0.2185)$
and $(B_0, \alpha)=(0.4229 \pm 0.1134, 4.2889 \pm 1.2173)$ respectively.
As we see from this figure both the form of anisotropic power spectrum are able to explain the lack of power for the multipole
range $l=2-10$.
For the multipole range $l=10-40$, our model fails to explain the observed lack of power.
Our theoretical model power spectrum also has some disagreement with the observed data at high-l which we neglect for the time being.
The contour plots for both the form of anisotropic power spectrum are shown in
Fig. \ref{fig:contour1} and \ref{fig:contour2}. The Fig. \ref{fig:contour1} and \ref{fig:contour2} says that our theoretical parameters
have a very poor correlation with each other.
If we see the best-fit parameters given in Table \ref{tab:para}, then the correction to the isotropic primordial power spectrum
due to the anisotropic power spectrum affect all the six $\Lambda$CDM parameters themselves. Out of these six $\Lambda$CDM parameters,
five parameters differ by a small quantity from PLANCK 2015 best-fit result whereas the $\tau$ parameter differs a lot.
Hence to explain the lack of power spectrum throughout the
observed multipole range $l=2-40$, our theoretical model is not so efficient.

\begin{table}
\bc
\begin{minipage}[]{100mm}
\caption[]{prior range used in parameter estimation analysis for power law\label{tab:prior1}}\end{minipage}
\setlength{\tabcolsep}{1pt}
\small
\begin{tabular}{ccc}
\hline\noalign{\smallskip}
Parameter Name& Symbol & Prior Ranges \\
Baryon Density & $\Omega_b{h}^2$ & [0.005, 0.1] \\
Cold Dark Matter Density & $\Omega_{c}{h}^2$ & [0.001, 0.99] \\
Angular size of Acoustic Horizon& $100\theta_{\textrm{MC}}$ & [0.5, 10.0] \\
Optical Depth & $\tau$ & [0.01, 0.8]  \\
Scalar Spectral Index & $n_s$  &  [0.8, 1.2] \\
Scalar Amplitude & ln($10^{10} A_s$) & [2, 4]\\
anisotropic spectral index   & $\alpha$  & [0, 0.8] \\
Anisotropic amplitude  & $B_0$  & 0.04 \\
\noalign{\smallskip}\hline
\end{tabular}
\ec
\end{table}

\begin{table}
\bc
\begin{minipage}[]{100mm}
\caption[]{prior range used in parameter estimation analysis for exponential power\label{tab:prior2}}\end{minipage}
\setlength{\tabcolsep}{1pt}
\small
\begin{tabular}{ccc}
\hline\noalign{\smallskip}
Parameter Name& Symbol & Prior Ranges \\
Baryon Density & $\Omega_b{h}^2$ & [0.005, 0.1] \\
Cold Dark Matter Density & $\Omega_{c}{h}^2$ & [0.001, 0.99] \\
Angular size of Acoustic Horizon& $100\theta_{\textrm{MC}}$ & [0.5, 10.0] \\
Optical Depth & $\tau$ & [0.01, 0.8]  \\
Scalar Spectral Index & $n_s$  &  [0.8, 1.2] \\
Scalar Amplitude & ln($10^{10} A_s$) & [2, 4]\\
anisotropic spectral index   & $\alpha$  & [0, 8]\\
Anisotropic amplitude  & $B_0$  & [-1, 1] \\
\noalign{\smallskip}\hline
\end{tabular}
\ec
\end{table}

\begin{table}
\bc
\begin{minipage}[]{100mm}
\caption[]{The best-fit parameter values with $1\sigma$ error obtained from MCMC analysis.
The first column represents the PLANCK 2015 best-fit $\Lambda$CDM parameter value,
the second and third column represents the parameter values for our theoretical model having power law
and the exponential form of the anisotropic power spectrum respectively\label{tab:para}}\end{minipage}
\setlength{\tabcolsep}{1pt}
\small
\begin{tabular}{cccc}
\hline\noalign{\smallskip}
Parameter & best-fit($\Lambda$CDM) & best-fit(with power law) & best-fit(with Exponential form)\\
$\Omega_b{h}^2$   & $0.02222 \pm 0.00023$ & $0.02037 \pm 0.00021$   & $0.01938 \pm 0.00031$\\
$\Omega_{c}{h}^2$ & $0.1197 \pm  0.0022$   & $0.1255   \pm 0.0025$    & $0.1430  \pm 0.0053$\\
$100\theta_{\textrm{MC}}$      & $1.04085 \pm  0.00047$  & $1.03934   \pm 0.00046$   & $1.03789  \pm 0.00059$\\
$\tau$            & $0.078  \pm  0.019$    & $0.057   \pm 0.022$     & $0.053   \pm 0.020$ \\
$n_s$             & $0.9655  \pm  0.0062$   & $0.9365   \pm 0.0081$     & $0.9794   \pm 0.0144$\\
ln($10^{10} A_s$)  & $3.098   \pm  0.036$    & $3.067     \pm 0.046$     & $3.042    \pm 0.037$\\
$\alpha$(model parameter)          &        &  $0.4556     \pm 0.2158$                     & $4.2889    \pm 1.2173$\\
$B_0$(model parameter)
            &      & $ 0.04 $ & $0.4229 \pm 0.1134$   \\
\noalign{\smallskip}\hline
\end{tabular}
\ec
\end{table}

\section{Conclusions}
\label{sec:con}
In this piece of work, we have analyzed direction dependent power spectrum obtained from
Finsler space-time. Here we have considered up to the second order correction of the primordial
power spectrum. The first order correction of the power spectrum produced the correlation
between the multipoles differ by $\Delta l=1$, whereas the second order correction produced the
correlation between the multipoles differs by $\Delta l=2$ in addition to the multipoles differ by $\Delta l=0$.
We found that the correlation between the multipoles differ by $\Delta l=0$ has a
contribution to the isotropic angular power spectrum $C^{TT}_l$. Here we have interested only on this correction
term of the isotropic angular power spectrum and studied its effect on the observed low-$l$ anomalies in the CMB data.
We have explicitly studied the lack of power in the low
multipole range $l\leq 40$.
We have parameterized the anisotropic power spectrum $B(k)$ of the quadrupolar modulation term and used in
CosmoMC software to determine best-fit model parameters using PLANCK likelihood code.
We have taken the power law as well as
an exponential form of the anisotropic power spectrum. For the power law form of the anisotropic
spectrum, we found that for $B_0=0.04$ and $\alpha = 0.4556 \pm 0.2158$, our model can able to explain the lack of power in the multipole
range $l=2-10$. Whereas to explain the lack of power
in the same multipole range,
the exponential form of the anisotropic power spectrum took $\alpha=4.2889 \pm 1.2173 $ and $B_0= 0.4229 \pm 0.1134$.
But for the multipole range $l=10-40$, our theoretical model approaches the $\Lambda$CDM result.
Hence we found that our theoretical model could not explain the lack of power for the observed range of multipoles ($l=2-40$) significantly.
This may indicate to a more complex form of the anisotropic model which could be able to explain all the low-$l$ anomalies successfully.

\normalem
\begin{acknowledgements}
We acknowledge the use of CosmoMc software for our analysis. This work has been funded by the National Natural Science Foundation of China under
Grant no. 11375203, 11675182 and 11690022.
\end{acknowledgements}

\bibliographystyle{raa}
\bibliography{ref-new}

\end{document}